\documentstyle[twocolumn,aps,epsfig]{revtex}
\begin{document}
\draft
\title{Physical Kinetics of Ferroelectric Hysteresis}
\author{S. Sivasubramanian and A. Widom}
\address{Physics Department, Northeastern University, Boston MA 02115}
\maketitle

\begin{abstract}
The physical kinetics of single domain ferroelectric materials  
are studied using Landau-Khalatnikov equation. 
The hysteretic curves are obtained numerically. 
The effective coercive electric field theoretically varies  
with the driving amplitude and frequency.
The effects of thermal noise are explored 
using the Fokker-Planck kinetic equation. 
The ferroelectric switching times are discussed and 
Quantum effects are briefly explored.
\end{abstract}  

\narrowtext

\section{Introduction} 

Ferroelectric materials are endowed with a thermodynamic 
spontaneous electric dipole moment per unit volume 
\begin{math}{\bf P}_s\end{math} at  
a minimum of the free energy per unit volume 
\begin{math}F({\bf P},T)\end{math}. 
Both experimentally
\cite{1,2,3,4,5,6,7,8,9,10,11,12,13,14,15,16,17,18}
and theoretically
\cite{19,20,21,22,23,24,25,26,27,28}, 
the polarization can be changed in 
direction via a dynamic {\em switching} process accompanied by hysteresis 
\cite{9,28,29,30,31,32,33,34,35,36,37,38,39,40,41,42,43,44,45,46,47}. 
Some of the partially understood experimental properties of 
ferroelectric domains \cite{27,48,49,50,51,52,53,54,55,56,57,58} 
include the following: 
(i) No unique coercive field is evident during the hysteretic 
process of switching. \cite{28,29,30,31,32,33,34,35,36,37,38,39} 
(ii) When an alternating electric field 
is applied, a hysteretic loop is observed only within a limited 
frequency band and a limited range of the amplitude-frequency 
product.\cite{9,43,44,45,46} 
(iii) Formation of multiple hysteresis loops \cite{47} are commonly 
observed.

From the thermodynamic viewpoint, one often assumes a phenomenological 
form for the free energy per unit volume 
\begin{math}F({\bf P},T)\end{math} and 
employs the thermodynamic law 
\begin{equation}
dF= -SdT+{\bf E}\cdot d{\bf P}.
\end{equation} 
The thermodynamic electric field is given by
\begin{equation}
{\bf E}=\left ({\partial F\over \partial {\bf P}} \right)_T 
\ \ \ {\rm (equilibrium)}.
\end{equation} 
The phenomenological free energies adequately describe the 
equilibrium experimental properties of ferroelectric materials.  

The dynamical hysteretic properties of the ferroelectric domains 
are less well studied from a theoretical viewpoint \cite{31,40,41,42}. 
In principle, a changing polarization 
is equivalent to a current density 
\begin{math}{\bf J}=(d{\bf P}/dt)\end{math}. 
One expects a current density to give rise to a non-equilibrium 
electric field 
\begin{math}{\bf E}_{dynamical} =\rho {\bf J} =\rho (d{\bf P}/dt)\end{math} 
where \begin{math}\rho \end{math} 
is an ``internal resistivity''. The above argument leads to the  
Landau-Khalatnikov dynamical equation of motion \cite{59,60,61,62}, 
\begin{equation}
{\bf E}=\left( {\partial F\over \partial {\bf P}}\right)_T
+\rho \left ({d{\bf P}\over dt} \right)
\ \ \ {\rm (non-equilibrium)}.
\end{equation}

The Landau-Khalatnikov Eq.(3) has been employed most often in the 
linearized form close to thermal equilibrium 
\begin{equation}
\delta {\bf E}=\left({1\over \chi }+
\rho {d\over dt} \right)\delta {\bf P},
\end{equation}
where \begin{math} \chi  \end{math} is the isothermal electric 
susceptibility. From the linearized  Landau-Khalatnikov Eq.(4), 
one may deduce the relaxation time 
\begin{equation}
\tau =\rho \chi.
\end{equation}
While the relaxation time \begin{math} \tau \end{math} can be 
employed to estimate the switching time, it is evident that hysteretic 
loops are inherently non-linear. Our purpose is to solve the non-linear 
Landau-Khalatnikov Eq.(3) with a view towards a deeper theoretical 
understanding of ferroelectric hysteresis. The solutions to be discussed 
in the work which follows have been obtained numerically. The 
qualitative features deduced from the non-linear Landau-Khalatnikov Eq.(3) 
are in satisfactory agreement with experiment.

In Sec. II, the Landau-Khalatnikov equation is considered for the case 
in which the driving electric field has the form
\begin{math} {\bf E}(t)={\bf E}_0\cos (\omega t)  \end{math}.
The numerically obtained hysteretic curves will be exhibited in 
Sec. III. In Sec. IV the effects of thermal noise are explored 
using the Fokker-Planck physical kinetic equation \cite{63} for the 
probability distribution. The ferroelectric switching time 
is computed as a function of the applied electric field. 
In Sec. V we present the Lagrangian form of the equations of motion 
and extend the theory to include second derivative terms in the 
differential equation of motion. The quantum mechanical consequences 
of the Lagrangian formulation of the physical kinetics are 
discussed in Sec.VI. Future prospects for theoretical 
extensions are discussed in the concluding Sec. VII.
       
\section{Landau-Khalatnikov Equation}

To measure Hysteretic cycles one applies an electric field  
to the ferroelectric sample 
\begin{math} {\bf E}(t)={\bf E}_0 \cos(\omega t)\end{math}. 
The Landau-Khalatnikov equation 
\begin{equation}
\rho \left({d{\bf P}\over dt}\right)+
\left({\partial F({\bf P},T)\over \partial {\bf P}}\right)
={\bf E}_0 \cos(\omega t),
\end{equation}
then determines how the polarization \begin{math} {\bf P}(t) \end{math} 
responds in time. For a stable hysteretic curve, one looks for 
a periodic solution 
\begin{equation}
{\bf P}\big(t+(2\pi /\omega )\big)={\bf P}(t)
\ \ \ ({\rm cyclic\  process}).
\end{equation}
If the polarization \begin{math} {\bf P}=(0,0,P) \end{math} and the 
applied electric field \begin{math} {\bf E}=(0,0,E) \end{math} 
are along an ``easy axis'', then Eq.(6) reads 
\begin{equation}
\rho \left({dP\over dt}\right)+
\left({\partial F(P,T)\over \partial P}\right)
=E_0 \cos(\omega t). 
\end{equation}
We choose for a model free energy, the Landau phenomenological form 
\begin{equation}
F(P,T)=\left({1\over 8\chi_0(T) P_s(T)^2}\right)
\left(P^2-P_s(T)^2\right)^2,
\end{equation}
where \begin{math} P_s(T)  \end{math}and 
\begin{math} \chi_0(T)  \end{math} are, respectively, the polarization 
and electric susceptibility with a zero static electric field. 

For the purpose of numerical integration of Eqs.(8) and (9), one may 
introduce the dimensionless quantities
\begin{equation}
\theta =\omega t,\ \ x=(P/P_s),\ \ {\rm and}\ \ y=(\chi_0 E/P_s).
\end{equation}
Also 
\begin{equation}
\eta =\left({1\over 2\chi_0\rho \omega }\right)
\ \ {\rm and} \ \ z_0=\left({E_0\over \rho \omega P_s}\right).
\end{equation}
Eqs.(8)-(11) now read 
\begin{equation}
\left({dx\over d\theta }\right)+\eta x(x^2-1)=z_0 \cos \theta .
\end{equation}
To calculate a hysteretic loop one must numerically solve Eq.(12) 
subject to an initial condition \begin{math} x(\theta =0) \end{math}. 
After integrating Eq.(12) over one period of \begin{math} 2\pi  \end{math}, 
one finds that 
\begin{equation}
x(\theta =2\pi )={\cal F}\big(x(\theta=0);\eta , z_0\big).
\end{equation}
The initial condition \begin{math} x(\theta =0)=x \end{math} 
for a hysteretic loop 
\begin{math} x(\theta +2\pi )=x(\theta ) \end{math}  
must be a fixed point of the function  
\begin{math} {\cal F} \end{math}; i.e. 
\begin{equation}
x={\cal F}(x;\eta ,z_0).
\end{equation} 
Such fixed points exist only for a limited region of values of 
the parameters \begin{math} (\eta ,z_0) \end{math}. Thus, only for 
finite ranges of frequency \begin{math} \omega \end{math} and 
driving electric field \begin{math} E_0 \end{math} in Eq.(8) will 
give rise to clean periodic hysteresis loops. For some regimes of 
frequency and driving electric field, the motion in the 
phase plane \begin{math} (E,P)  \end{math} will be chaotic. 
The limited range of parameters \begin{math} (\eta ,z_0) \end{math}  
for observing hysteretic periodic motions in the phase plane is in 
agreement with well known features observed in ferroelectric  
experiments.

\section{Numerical Results}

Employing Eqs.(1) and (9), we find that the thermodynamic 
electric field implied by the model under consideration, 
\begin{equation}
E=\left({\partial F\over \partial P}\right)_T
\ \ \ {\rm (equilibrium)},
\end{equation}
obeys 
\begin{equation}
E=\left({P\over 2\chi_0 P_s^2}\right)(P^2-P_s^2)
\ \ \ \ ({\rm thermal}).
\end{equation}
The purely thermodynamic Eq.(16) gives rise to hysteretic behavior 
at a switching polarization of 
\begin{equation}
P_{switch}=\left({P_s\over \sqrt{3}}\right),
\end{equation}
which occurs at a coercive electric field of 
\begin{equation}
E_c=\left({P_s\over 3\chi_0 \sqrt{3}}\right)
=\left({P_{switch}\over 3\chi_0}\right).
\end{equation}

\begin{figure}[htbp]
\begin{center}
\mbox{\epsfig{file=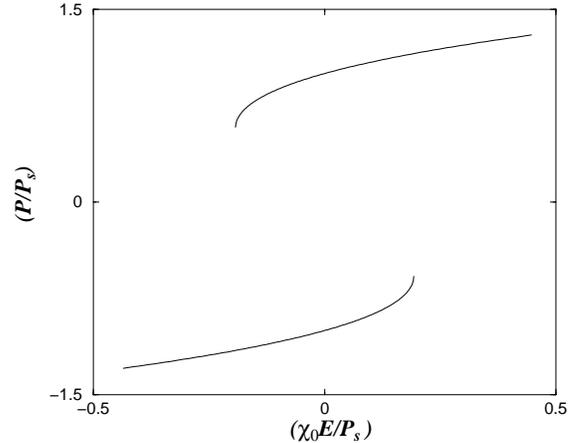,height=60mm}}
\caption{Shown is a metastable thermal switching hysteresis curve 
calculated from Eq.(16). If the applied field starts negatively at 
$E<-E_c$ and the applied field is subsequently quasi-statically increased, 
then the polarization follows the lower curve and jumps to the 
upper curve when the coercive field $E_c$ is exceeded.}
\label{fig1}
\end{center}
\end{figure}

\begin{figure}[htbp]
\begin{center}
\mbox{\epsfig{file=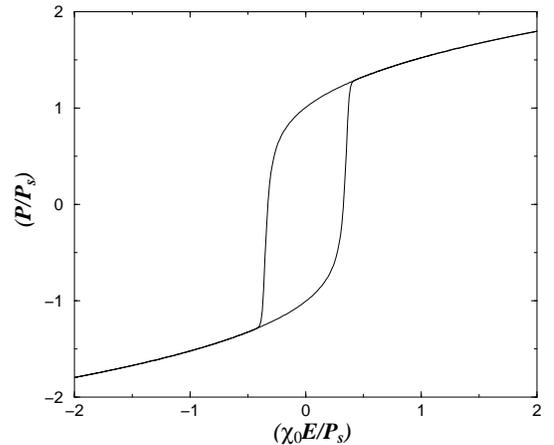,height=60mm}}
\caption{The above plot is a numerically simulated hysteric 
loop calculated from the Landau-Khalatnikov model. 
The parameters chosen were $\eta =50.0$ 
and $z_0=200.0$.}
\label{fig2}
\end{center}
\end{figure}
 
When a cyclic process is numerically simulated for the case of 
finite frequency as in FIG.2, the resulting hysteretic loop looks 
qualitatively similar to the thermodynamic FIG.1.

However, there are quantitative features which depend 
on the magnitude of the frequency and the amplitude of the 
applied oscillating electric field. For example the dynamical 
``coercive field'' deduced from the numerical simulations at finite 
frequency is quite different from the static thermodynamic 
``coercive field''. Furthermore, the ``shape'' of the hysteretic loop 
depends strongly on the product of the frequency and the internal 
resistivity via \begin{math} 2\eta=(\chi_0\rho \omega )^{-1} \end{math}. 
This point is illustrated in FIG.3. 
\begin{figure}[htbp]
\begin{center}
\mbox{\epsfig{file=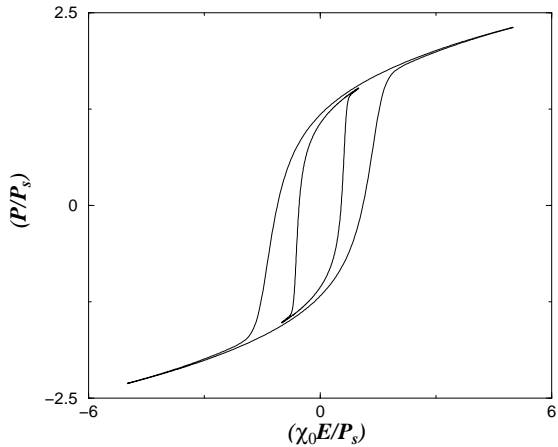,height=60mm}}
\caption{Two {\it different} numerically simulated hysteretic 
loops are shown. The smaller (inner) loop 
corresponds to the values $\eta =5.0$ and $z_0=10.0$. The larger 
(outer) loop corresponds to the values $\eta =5.0$ and $z_0=50.0$. 
The two loops correspond to the same frequency. However, the larger 
loop is induced by the larger amplitude of the applied electric field.}
\label{fig3}
\end{center}
\end{figure}
  
It is evident, that the dynamically induced coercive field (for fixed 
frequency) depends on the amplitude of the applied electric field. Thus, 
the large (outer) hysteretic loop in FIG.3 has a much larger dynamic 
coercive field than does the small (inner) hysteretic loop. It is not 
in general possible to deduce the thermodynamic coercive field of 
Eq.(18) by the observation of a single dynamical hysteretic loop. 
From an experimental point of view, this ambiguity in determining the 
thermodynamic coercive field is a long standing problem.

\section{Thermal Noise}

The existence of an internal resistivity 
\begin{math} \rho  \end{math} in the  Landau-Khalatnikov Eq.(3) 
requires a thermally fluctuating electric field   
\begin{math} \Delta {\bf E}(t) \end{math}; i.e. Eq.(3) with the 
inclusion of electric field noise reads 
\begin{equation}  
\rho \left ({d{\bf P}\over dt} \right) ={\bf E}-
\left( {\partial F\over \partial {\bf P}}\right)_T+ \Delta {\bf E}.
\end{equation}
We have included ``white noise'' field fluctuations obeying   
\begin{equation}
\overline{\Delta {\bf E}(t)\Delta {\bf E}(t^\prime )}=
\left({k_BT\over \Omega }\right)\rho {\bf 1}\delta (t-t^\prime ).
\end{equation}
If \begin{math} W({\bf P},t)d^3{\bf P} \end{math} denotes the 
probability for \begin{math} {\bf P}\in d^3{\bf P} \end{math}, 
then the electric field noise in Eqs.(19) and (20) give rise 
to the Fokker-Planck equation 
\begin{equation}
\rho {\partial W\over \partial t}=
{\partial \over \partial {\bf P}}\cdot 
\left\{
\left({\partial G\over \partial {\bf P}}\right)_{{\bf E},T}W+
\left({k_B T \over \Omega }\right)
{\partial W\over \partial {\bf P}}\right\},
\end{equation} 
where 
\begin{equation}
G({\bf P},{\bf E},T)=F({\bf P},T)-{\bf E\cdot P}. 
\end{equation}
In thermal equilibrium, the probability is determined by 
\begin{equation}
W_{eq}({\bf P})=
\exp\left({\Omega 
\big(A({\bf E},T)-G({\bf P},{\bf E},T)\big)\over k_BT}\right),
\end{equation}
where \begin{math} A({\bf E},T) \end{math} is determined by the 
total probability normalization 
\begin{equation}
\int W_{eq}({\bf P})d^3{\bf P}=1.
\end{equation}
If the polarization \begin{math} {\bf P}=(0,0,P) \end{math} and the 
applied electric field \begin{math} {\bf E}=(0,0,E) \end{math} 
are along an ``easy axis'', then it is sufficient to use a 
one-dimensional probability \begin{math} W(P,t)dP  \end{math} for 
\begin{math} P\in dP  \end{math} which obeys a one-dimensional 
Fokker-Planck equation
\begin{equation}
\rho {\partial W\over \partial t}=
{\partial \over \partial P}
\left\{
\left({\partial G\over \partial P}\right)_{E,T}W\right\}
+
\left({k_B T \over \Omega }\right)
{\partial^2 W\over \partial P^2}\ .
\end{equation}

The ``jump time'' associated with Eq.(25) may be computed in a 
well known manner. Suppose that the polarization at time zero is 
given by \begin{math} P(t=0)=P_i \end{math}. With a final polarization 
\begin{math} P_f>P_i \end{math}, let 
\begin{math} \tau [P_f,P_i] \end{math} 
be the mean ``first passage time'' 
for the polarization to move from a region 
\begin{math} P<P_f \end{math} to the complimentary region 
\begin{math} P>P_f \end{math}. In detail, let 
\begin{math} t^* \end{math} be the smallest positive time for which 
\begin{math} P(t^*)=P_f \end{math}. The mean first passage time 
is then \begin{math} \tau=\overline{t^*} \end{math}.

The closed form expression for the mean first passage time is
well known to be
$$  
\tau [P_f,P_i]=\left({\Omega \rho \over k_B T}\right)\times 
$$
\begin{equation} 
\int_{P_i}^{P_f}\int_{-\infty }^P 
e^{\Omega \big(G(P,E,T)-G(P^\prime ,E,T)\big)/k_B T}dP^\prime dP.
\end{equation}
In principle, it is possible to solve the equation 
\begin{equation}
t=\tau [P_f,P_i]
\end{equation} 
for the final polarization 
\begin{equation}
P_f=P(t;P_i).
\end{equation}
Eqs.(26), (27) and (28) imply 
\begin{equation}
\rho \left({\partial P\over \partial t}\right)+
\left({\partial F\over \partial P}\right)_T=E+\tilde{E},
\end{equation}
where 
\begin{equation}
\left({\partial P\over \partial t}\right)^2\tilde{E}=-
\left({k_BT \over \Omega }\right)
\left({\partial^2 P\over \partial t^2}\right).
\end{equation}

In the limit of large single domains 
\begin{math} \Omega \to \infty  \end{math}, we have 
\begin{math} \tilde{E} \to 0  \end{math} so that Eq.(29) becomes 
equivalent to the Landau-Khalatnikov Eq.(3). When 
\begin{math} |\tilde{E}|<<|E| \end{math}, then the  
Landau-Khalatnikov equation is sufficiently accurate for computing 
first passage times. However, very near critical temperatures and or 
critical electric (coercive) fields, a finite domain can switch due to 
thermal noise. The integral in Eq.(26) for the first passage time 
must then be evaluated. 

\section{Lagrangian Formulation}

In order to understand the extended Landau-Khalatnikov-Tani equation 
\cite{64} for ferroelectric single domains, let us first review 
the extended Landau-Lifshitz-Gilbert equations \cite{65,66} for 
ferromagnetic single domains. For the ferromagnetic case, 
there exists a  free energy per unit volume 
\begin{math}F({\bf M},T)\end{math} which obeys the thermodynamic law 
\begin{equation}
dF= -SdT+{\bf H}\cdot d{\bf M}.
\end{equation} 
The thermodynamic magnetic intensity is given by
\begin{equation}
{\bf H}=\left ({\partial F\over \partial {\bf M}} \right)_T 
\ \ \ {\rm (equilibrium)}.
\end{equation} 
For non-equilibrium situations in an applied magnetic intensity 
\begin{math} {\bf H} \end{math}, there is an effective field 
\begin{equation}
{\bf H}_{eff}={\bf H}-
\left({\partial F\over \partial {\bf M}}\right)
-\rho \left({\partial {\bf M}\over \partial t}\right),
\end{equation}
The equation of motion for the magnetization then reads 
\begin{equation}
\left({\partial {\bf M}\over \partial t}\right)=
\gamma {\bf M\times H}_{eff},
\end{equation}
where \begin{math} \gamma \end{math} is the gyromagnetic ratio.

For the case of a Ferroelectric, Eqs.(1) and (2) replace 
Eqs.(31) and (32), and the effective magnetic intensity 
Eq.(33) is replaced by the effective electric field 
\begin{equation}
{\bf E}_{eff}={\bf E}-
\left({\partial F\over \partial {\bf P}}\right)_T
-\rho \left({\partial {\bf P}\over \partial t}\right).
\end{equation}
While in the Landau-Khalatnikov theory, the effective field 
\begin{math} {\bf E}_{eff}  \end{math} is set to zero, the Tani 
extension to the Landau-Khalatnikov theory includes an inertial 
term of the form 
\begin{equation}
4\pi \left({\partial ^2 {\bf P}\over \partial t^2}\right)=
\omega_p^2 {\bf E}_{eff}.
\end{equation}
Let us consider Eq.(36) in more detail. 
The dipole moment of \begin{math} N  \end{math} charged 
particles in a volume \begin{math} \Omega \end{math} is 
defined as 
\begin{equation}
\Omega {\bf P}=\sum_{j=1}^N e_j {\bf r}_j,
\end{equation}
where \begin{math} {\bf r}_j \end{math} is the position of the 
\begin{math} j^{th} \end{math} particle with charge 
\begin{math} e_j \end{math}. From the equation of motion for  
the \begin{math} j^{th} \end{math} particle with mass 
\begin{math} m_j \end{math}, 
\begin{equation}
m_j\ddot{\bf r}_j=e_j {\bf E}_{eff} 
\end{equation}
and Eq.(37) one easily obtains Eq.(36) wherein 
\begin{math} \omega_p  \end{math} can be identified with the 
plasma frequency 
\begin{equation}
\omega_p^2=\left({4\pi \over \Omega }\right)\sum_{j=1}^N 
\left({e_j^2\over m_j}\right).
\end{equation}
Eqs.(35) and (36) yield the Landau-Khalatnikov-Tani theory for 
ferroelectric domains; i.e. 
\begin{equation}
\left({4\pi \over \omega_p^2}\right)
\left({\partial^2 {\bf P}\over \partial t^2 }\right)+
\rho \left({\partial {\bf P}\over \partial t}\right)+
\left({\partial F\over \partial {\bf P}}\right)_T={\bf E}.
\end{equation}
Eq.(40) may be written in terms of the effective Lagrangian 
\begin{math} L=\Omega {\cal L} \end{math};
\begin{equation}
{\cal L}=\left({2\pi \over \omega_p^2}\right)
\left({\partial {\bf P}\over \partial t}\right)^2
-F({\bf P},T)+{\bf E\cdot P}
\end{equation}
and the Rayleigh dissipation function 
\begin{math} R=\Omega {\cal R} \end{math};
\begin{equation}
2{\cal R}=\rho \left({\partial {\bf P}\over \partial t}\right)^2. 
\end{equation}
We have 
\begin{equation}
{\partial\over \partial t}
\left({\partial {\cal L}\over \partial 
(\partial {\bf P}/\partial t)}\right)=
\left({\partial {\cal L}\over \partial {\bf P}}\right)
-\left({\partial {\cal R}\over \partial 
(\partial {\bf P}/\partial t)}\right).
\end{equation}
The canonical ``momentum'' \begin{math} {\bf \Pi } \end{math} 
conjugate to the polarization \begin{math} {\bf P} \end{math}
may be defined by 
\begin{equation}
{\bf \Pi }=\Omega \left({\partial {\cal L}\over \partial 
(\partial {\bf P}/\partial t)}\right)=
\left({4\pi \Omega \over \omega_p^2 }\right)
\left({\partial {\bf P}\over \partial t}\right).
\end{equation}
Employing the Lagrangian viewpoint, one may quantize the single 
ferroelectric domain theory in the usual manner.

\section{Quantum Kinetics}

In the fully quantum mechanical theory one expects the canonical 
commutation relation 
\begin{equation}
\left[{\bf \Pi },{\bf P}\right]=-i\hbar {\bf 1} 
\end{equation}
which may also be proved employing a microscopic view. From 
Eq.(37) and (44) it follows 
(with \begin{math} \dot{\bf r}_j={\bf v}_j \end{math}) that 
\begin{equation}
{\bf \Pi }=\left({4\pi \over \omega_p^2}\right)
\sum_{j=1}^N e_j {\bf v}_j.
\end{equation}
From the microscopic commutation relation 
\begin{equation}
\left[{\bf v }_i,{\bf r}_j\right]=
-\left({i\hbar \delta_{ij}\over m_j}\right){\bf 1}, 
\end{equation}
together with Eqs.(37), (39) and (46) follows the macroscopic 
quantum mechanical commutator in Eq.(45). One representation of the 
quantum mechanical operator \begin{math}{\bf \Pi }\end{math} is 
defined by 
\begin{equation}
{\bf \Pi }=
-i\hbar \left({\partial \over \partial {\bf P}}\right).
\end{equation}
Forming the Hamiltonian from the Lagrangian in the 
standard canonical formalism, 
\begin{equation}
\Omega {\cal H}_{eff}={\bf \Pi \cdot P}-\Omega {\cal L},
\end{equation}
yields the effective Hamiltonian for a ferroelectric domain  
\begin{equation}
\Omega {\cal H}_{eff}=
\left({\omega_p^2\over 8\pi \Omega}\right){\bf \Pi }^2 
+\Omega \left\{F({\bf P},T)-{\bf E\cdot P}\right\}.
\end{equation}
where Eqs.(41), (44) and (49) have been invoked. 
Eqs.(48) and (50) describe a single domain ferroelectric grain 
in macroscopic quantum mechanical form.

From Eqs.(9), (48) and (50) one may compute the quasi-classical 
approximation for the quantum tunneling switching time. At zero 
electric field \begin{math} {\bf E}=0 \end{math}, the quantum 
transition rate per unit time to switch polarizations (say  
\begin{math} {\bf P}_s\to -{\bf P}_s \end{math}) is given by 
\begin{equation}
\left({1\over \tau_0 }\right)_{quantum}
\approx \gamma_0 \exp(-B_0)
\end{equation}
with a barrier factor given by 
\begin{equation}
B_0={8\over 3}\sqrt{\pi \over \chi_0}
\left(\Omega P_s^2\over \hbar \omega_p\right),
\end{equation}
and an attempt frequency of 
\begin{equation}
\gamma_0=\left({2\omega_0\over \pi}\right)
\sqrt{2\Omega U_0\over \hbar \omega_0}.
\end{equation}
In Eq.(53), the Barrier energy per unit volume is 
\begin{equation}
U_0=\left({P_s^2\over 8\chi_0}\right)
\end{equation}
and 
\begin{equation}
\omega_0^2=\left({\omega_p^2\over 4\pi \chi_0}\right).
\end{equation}
Thus
\begin{equation}
B_0={32\over 3}\left({\Omega U_0\over \hbar \omega_0}\right).
\end{equation}
Quantum tunneling through the barrier will dominate classical 
thermal activation over the barrier if 
\begin{equation}
\exp(-B_0)>>\exp(-\Omega U_0/k_B T).
\end{equation}
Thus for temperatures small on the scale of 
\begin{equation}
T^*=\left({\Omega U_0\over k_B B_0}\right)
=\left(3\hbar \omega_0\over 32 k_B\right),
\end{equation}
quantum tunneling will dominate classical thermal activation. 
For typical laboratory samples \begin{math} T<<T^* \end{math} 
so that the life-time of domains for very small volume domains 
is dominated by quantum mechanics. Nevertheless, for large volume 
domains there is stability and the kinetics are well described 
by the Landau-Khalatnikov kinetic equation. 

\section{Conclusion}

We have shown that physical kinetics, via the Landau-Khalatnikov 
equation, adequately describes laboratory hysteretic behavior
for large volume single domain ferroelectric materials.  
The important physical quantities include the amplitude 
and frequency dependence of the effective coercive electric field. 
The formalism describing thermal noise as well as 
quantum noise have been explored. Both forms of noise 
are unimportant for large volume ferroelectric domains. 
In the small domain volume limit, quantum noise effects dominate 
the classical thermal noise effects at laboratory temperatures 
substantially below the critical temperature. 

The technical importance of ferroelectric switching times 
for small ferroelectric domain volumes is well documented 
\cite{1,2,3,4,5,6,7,8,9,10,11,12,13,14,15,16,17,18}. 
The computation of switching times requires 
mesoscopic quantum mechanical models. Thus far, the theory 
of such quantum switches is in its infancy.

\end{document}